\documentclass[preprint2]{aastex}

\usepackage[normalem]{ulem}

\newcommand{\bq}{\begin{equation}}
\newcommand{\eq}{\end{equation}}
\newcommand{\bqn}{\begin{eqnarray}}
\newcommand{\eqn}{\end{eqnarray}}
\newcommand{\dd}{\mbox{\rm d}}

\newcommand{\Msol}{\mathrm{M}_\mathrm{\odot}}
\newcommand{\Lsol}{\mathrm{L}_\mathrm{\odot}}

\shorttitle{Accretion rates  of stars on Super-massive Black Holes}
\shortauthors{A. Just et al.}

\begin{document}


\title{Enhanced accretion rates of stars on Super-massive Black Holes
 by star-disk interactions in galactic nuclei}


\author{Andreas Just\altaffilmark{1},
Denis Yurin\altaffilmark{2,1}, 
Maxim Makukov\altaffilmark{2,1},
Peter Berczik\altaffilmark{3,1,5},
Chingis Omarov\altaffilmark{2,1},
Rainer Spurzem\altaffilmark{3,1,4},
and Emanuel Y. Vilkoviskij\altaffilmark{2}} 
\affil{$^1$Universit\"at Heidelberg, Zentrum f\"ur Astronomie, Astronomisches Rechen-Institut, 
M{\"o}nchhof-Stra{\ss}e 12-14, 69120 Heidelberg, Germany
\email{just@ari.uni-heidelberg.de}
$^2$Fesenkov Astrophysical
Institute, Observatory 23, 050020 Almaty, Kazakhstan\\
$^3$National Astronomical Observatories of China, Chinese Academy of Sciences,
20A Datun Rd., Chaoyang District, Beijing 100012, China\\
$^4$Kavli Institute for Astronomy and Astrophysics, Peking University, China\\
$^5$Main Astronomical Observatory, 
National Academy of Sciences of Ukraine,
MAO/NASU, 27 Akademika Zabolotnoho St.
03680 Kyiv, Ukraine}


\begin{abstract}
We investigate the dynamical interaction of a central star cluster
surrounding a super-massive black hole and a central
accretion disk. The { dissipative force acting on stars in the disk} leads to an enhanced mass flow
towards the super-massive black hole and to an asymmetry in the phase space distribution
due to the rotating accretion disk. 
The accretion disk is considered as a
stationary Keplerian rotating disk, which is vertically extended in
order to employ a fully self-consistent treatment of stellar dynamics
including the dissipative force originating from star-gas ram pressure
effects. The stellar system is treated with a direct high-accuracy
$N$-body integration code. A star-by-star representation, desirable in $N$-body
simulations, cannot
be extended to real particle numbers yet. Hence, 
we carefully discuss the scaling behavior of our model with
regard to particle number and tidal accretion radius. 
The main idea is to find a family of models for which the
ratio of two-body relaxation time and dissipation time (for kinetic
energy of stellar orbits) is constant, which then allows us to extrapolate
our results to real parameters of galactic nuclei.
Our model is derived from basic physical principles and as such
it provides insight into the role of physical processes in galactic nuclei, but
it
should be regarded as a first step towards more realistic and more comprehensive
simulations. Nevertheless, the following conclusions appear to be robust:
the star accretion rate onto the accretion disk and subsequently onto
the super-massive black hole is enhanced by a significant factor compared to 
{ purely stellar dynamical} systems
neglecting the disk. This process leads to enhanced fueling of
central disks in active galactic nuclei and to an enhanced rate of
tidal stellar disruptions. 
Such disruptions may produce electromagnetic counterparts in form of observable
X-ray flares. Our models improve predictions for their rates in 
quiescent galactic nuclei. We do not yet model direct stellar collisions
in the gravitational potential well of the black hole, which could further
enhance the growth rate of the black hole. Our models are relevant for
quiescent galactic nuclei, because all our mass accretion rates would give
rise to luminosities 
%change by Andreas
much smaller than the
%smaller or at most just equal to the 
Eddington luminosity. To reach
%super
Eddington luminosities, outflows and feedback
as in the most active QSO's other scenarios are needed, such as
gas accretion after galaxy mergers.
%{ which will be important to predict electromagnetic
%counterparts of black hole activity in galactic nuclei}.
However, for AGNs close to the Eddington limit this process may not serve as the dominant accretion process due to the long timescale.
\end{abstract}

%% example has been keyed in \apj style. See the instructions to authors

\keywords{accretion disks - galaxies, galactic nuclei -
celestial mechanics - stellar dynamics, N-body simulations}



\section{Introduction}\label{sec-intro}

Kinematic and photometric data of galactic nuclei have revealed that
super-massive central black holes (SMBHs) are ubiquitous in most
galaxies 
(see e.g. \citealt{KormendyR:95} for a review). 
Detailed information on their photometric profiles and shapes of
spectral 
line profiles allow us, in
certain limits, to deduce the true shape of the distribution function
in phase space of such systems. The distribution function determines
the rate at 
which stars will come close to the
SMBH, be tidally disrupted or destroyed by direct collisions and
eventually accreted
onto the black hole \citep{FrankR:76,BahcallW:76}. Quasars, however,
which are the most luminous witnesses of accretion activity onto SMBHs
are observed already in the young universe at redshifts of $z>5$. It
is difficult 
to explain how black holes can grow so quickly to the 
observed high masses
($10^6-10^9\Msol$) by pure star accretion. It has been therefore argued that
massive seed black holes form already by dissipative and viscous
collapse, 
possibly accompanied by the formation of massive stars and
their coalescence, at the time of galaxy formation \citep{Colgate:67,
  SpitzerS:66, SpitzerS:67, Sanders:70, Rees:84}. In the hierarchical
picture of galaxy formation, the most massive dark halos with small
angular momentum can 
account for the early formation of the most massive black holes. 
The SMBH forms 
by a centrally focused collapse entering a phase of a dense
super-massive gaseous object which is supported by star-gas
interactions 
\citep{BisnovatyiS:72,Vilkoviskij:75,Hara:78,Langbeinetal:90}.
The interaction of a compact stellar cluster with a massive central
object in the form of a superstar or SMBH was considered by
\citet{Vilkoviskij:75,Hills:75,Hara:78}. The evolution of the dense
non-rotating stellar cluster 
was studied by \citet{SpitzerS:66, Bisnovatyi:78} among others, 
and the evolution of the gas sphere was considered by \citet{Langbeinetal:90}.

These studies have commonly neglected angular momentum, which should not be
neglected during mergers nor in the intrinsic structure of galaxies. 
Spectra of active galactic nuclei suggest that there 
should be gas in the form of a massive central accretion disk (AD) in which all 
interstellar matter settles before it may be accreted. The origin 
of the AD could be inflowing cool gas during mergers 
and/or debris by direct stellar collisions or stellar evolutionary processes, 
depending on the evolutionary state of the host galaxy. \citet{Artymowiczetal:93} 
provide a theoretical model framework in which star-gas 
interactions and the build-up of massive stars generate a massive AD around the central SMBH.

Recent work shows that gaseous disks in galactic centers around SMBHs are important for 
the dynamics and the morphology of central regions of galaxies and for the evolution 
of single or multiple central black holes in many ways. For example
\citet{Cuadra:09,Callegari:09,Callegari:10} 
study the role of small scale disks for the acceleration of binary black hole mergers, 
through dissipation of kinetic energy in the disk,
after a galactic merger. \citet{Baruteau:10} discuss the hardening of stellar binaries 
in circumnuclear disks and their subsequent interactions with central black holes, which may lead to
high velocity stellar escapers. 

\citet{ShakuraS:73} developed                                        
a model for a stationary AD, which has been the basis of many investigations
since then.
Close to the inner boundary of the AD, at a few Schwarzschild radii $r_\mathrm{S}$,
general relativistic effects must be taken into account. 
\citet{NovikovT:73} extended the standard disc model in that regime.

If the inner part of the AD reaches a critical surface density the interaction
of orbits with
the gas (or more correctly energy and momentum transfer of stars due to ram
pressure effects,
{ henceforth denoted as dissipation}) cannot be neglected anymore. Stellar
interactions with accretion disks were also considered by
\citet{VilkoviskijB:82, Vilkoviskij:83, Syeretal:91}. More detailed investigations of the 
stellar orbits, crossing accretion disks were presented 
in \citet{VokrouhlickyK:98} and in \citet{SubrK:99}. 
\citet{SubrK:99} and \citet{Subr:03} assumed an infinitely thin disk interacting
with stars; they found that
the interaction between the disk and the stars (star-gas { dissipation}) will 
deplete counter-rotating stars, create a flattening of the large-scale structure 
of the system and initiate anisotropies (i.e. changes of the eccentricity distributions). 
These studies did not take into account any feedback onto the disk or any
finite-thickness 
effects of the disk. They tried to find a stationary state in which the
transport 
of stars into the central galactic region is balanced by removal of stars from
the central disk. Evolution timescales and initial conditions to reach this
equilibrium was missing. 

Around SMBHs the AD may extend to the parsec scale, orders of magnitude larger than $r_\mathrm{S}$. 
\citet{Rauch:95,Rauch:99} studied in detail the impact of the AD on the orbits 
and the distribution of stars with test particle simulations. Their work includes relativistic 
effects, even for the case of rotating Kerr black holes. They find that the orbital 
inclination with respect to the AD declines quickly as soon as the { dissipative} 
force becomes effective. Additionally they find a steepening of the central
density cusp, due 
to the combined effect of relativistic orbit migration and stellar collisions. 

The semi-analytic model by \citet{VilkoviskiC:02} raises the point that two-body relaxation of the
stars within and near the disk will tend to elevate trapped stars again out of the disk, 
and that the competition between relaxation and { dissipation} will define a 
stationary state of the system, with some well-defined stationary flux of stars 
going down to the black hole. \citet{VilkoviskiC:02}
compared the star-star two-body interactions with the star-disk interactions and concluded 
that the latter is stronger in the inner parts of the accretion disk and vice versa 
in the outer parts based on
nearly circular orbits of the stars. They derived analytic approximations for the effective 
inclination of stellar orbits, where the inflow to the SMBH takes place. 
They also derived a critical radius inside which 
direct stellar collisions must be taken into account.

All of the mentioned papers so far have both strong and weak points. \citet{Rauch:99} 
is the only paper to combine general relativity effects and stellar collisions, but
considered only the
central region where the (spherically symmetric) potential is dominated by the super-massive 
black hole. They used an approximate model of stellar dynamics based on a Monte
Carlo technique, which requires a spherically symmetric central star cluster. In
\citet{Rauch:95} 
they combined relativistic effects with star-disk interactions, but 
the disk is assumed to be infinitely thin. The infinitely thin disk
approximation was also used in other work by
\citet{SubrK:99,Subr:03,VilkoviskiC:02}).

Following the ideas of \citet{VilkoviskiC:02} we investigate by self-consistent
direct $N$-body simulations, the interaction of a central compact stellar
cluster (CSC) with the AD in active galactic nuclei. We focus on the mutual
interplay of two-body relaxation and the depletion of stars by the {
dissipative} force in the AD as a secular long-term evolution of the stellar
mass distribution and the velocity distribution function of the CSC. In this
paper we report results on a new model of star-disk interactions
in galactic nuclei. Our focus lies on the correct and accurate representation
of the stellar orbital motion crossing the disk, by implementing a disk with its density distribution in a full three-dimensional $N$-body simulation. We have added the force and force time derivative
to the standard Hermite scheme (see below) as a function of local disk density
and velocity in three dimensions. This is the most general approach, and later
it will allow us to
include evolving models of the AD and appropriately model the mass and energy
transfer between the AD and CSC. 

Our first results concern the enhanced accretion rate onto the central SMBH due
to the interaction with the AD, in the regime where the relaxation timescale is
comparable to the dissipation timescale. This is, to our knowledge, the first
approach to study the competition between relaxation and star-gas { dissipation}
in a direct simulation. Since the direct simulations are not yet able to reach
realistic particle numbers and spatial resolution, we will perform a careful
scaling analysis to show in which way our numerical simulations have to be
interpreted for real astrophysical galactic nuclei. 

This paper is organized as follows: in Sect.~\ref{sec-physics} we describe the
accretion disc and the { dissipative} force in detail, Sect.~\ref{sec-model}
gives the numerical realizations of the system, in Sect.~\ref{sec-result} the
results are discussed and in Sect.~\ref{sec-conclusion} a summary and
conclusions are presented.

In follow-up papers we will discuss the dependence of the { dissipation} on the
orbital parameters and the phase space evolution of the cusp stars in detail and
take into account the feedback of the star-gas interaction on the AD properties.
Detailed studies of migration of stars, binaries, and black holes inside the
disk towards the center, and its observational consequences will also be
included in future work.
 
\section{Physics of the accretion disk and the star-gas interactions} \label{sec-physics}

We consider an AGN model consisting of three subsystems: i) a
compact stellar cluster (CSC) with mass $M_\mathrm{cl}$ describing the
inner part of the galactic center. It is spherically symmetric,
non-rotating, and in dynamical equilibrium. ii) An accretion disk
(AD) with mass $M_\mathrm{d}=\mu_\mathrm{d}M_\mathrm{cl}$. It is vertically extended,
non-evolving, and has a Keplerian rotation curve. iii) A central
super massive black hole (SMBH) with mass $M_\mathrm{bh}=\mu_\mathrm{bh}M_\mathrm{cl}$. The
motion of each star $m_\mathrm{i}$ of the CSC is determined by the
mutual gravitational interaction of the stars, the gravitational
force of the SMBH, and by a dissipative force $\vec F_\mathrm{d}$
from the AD. The equation of motion is given by 
\bq 
\ddot{ \vec r_\mathrm{i}}=-\sum_\mathrm{i\not= j}\frac{Gm_\mathrm{j}\vec r_\mathrm{ij}}{r_\mathrm{ij}^3} -
\frac{GM_\mathrm{bh}\vec r_\mathrm{i}}{r_\mathrm{i}^3}+ \frac{\vec F_\mathrm{d}}{m_\mathrm{i}} \quad, 
\eq 
where $\vec r_\mathrm{ij}=\vec r_\mathrm{i}-\vec r_\mathrm{j}$ with $\vec r_\mathrm{i}$,
$\vec r_\mathrm{j}$ the positions of stars $i$ and $j$, respectively. Since
the AD has a small mass compared to the black hole and
the enclosed stellar cluster, we neglect the gravitating effect
of the disk on the system. Numerical details for computing the forces are given
in Sect. \ref{sec-model}.

\subsection{The accretion disk} \label{sub-disk}

We are interested in the dynamical action of the AD on the stellar
component. Therefore we adopt a 3-dimensional axisymmetric
stationary disk model, which is differentially rotating with the
local circular velocity. For the inner structure and evolution of
such disks see the review of \citet{ParkO:01}, and for
the basic physics of accretion disks see e.g.
\citet{Franketal:02}. A widely used model for the AD is the
model of \citet{ShakuraS:73}, with the radial scaling described 
in detail in \citet[hereafter NT]{NovikovT:73}, which was also used by
\citet{Rauch:95,VilkoviskiC:02}. 

For the radial profile of the AD the surface density $\Sigma$ is given by
\bq
\Sigma(R)=\Sigma_\mathrm{d}\left(\frac{R}{R_\mathrm{d}}\right)^{-\alpha}
\quad \mbox{with}\quad \alpha=3/4 , \label{eq-Sigma}
\eq
where $R^2=x^2+y^2$, $R_\mathrm{d}$ is the cut-off
radius of the disk and $\Sigma_\mathrm{d}$ is the surface density at the cut-off radius.
The value $\alpha=3/4$ corresponds to the outer disk range of NT.
 The mass $M_\mathrm{d}$ of the disk is 
\bq 
M_\mathrm{d}=2\pi \int_0^{R_\mathrm{d}}\Sigma(R)R\dd
R=\frac{2\pi}{2-\alpha}\Sigma_\mathrm{d}R_\mathrm{d}^2. \label{eq-Md} 
\eq 
For the numerical integration using the $4^{th}$order Hermite scheme we need a smooth force in space. 
Therefore we introduce a continuous but steep outer cutoff by
\bq
\Sigma(R)=\Sigma_\mathrm{d}\left(\frac{R}{R_\mathrm{d}}\right)^{-\alpha}
\exp\left[-\beta_\mathrm{s}\left(\frac{R}{R_\mathrm{d}}\right)^s\right],
\label{eq-sigcut}
\eq 
In order to retain exactly
Eq. \ref{eq-Md} for the total disk mass (with the integral now up to $\infty$)
we choose
\bq
\beta_\mathrm{s}=\left[\Gamma\left(1+\frac{2-\alpha}{s}\right)\right]^{s/(2-\alpha)}
\eq
with the Gamma-function $\Gamma(x)$. For $s\to \infty$ the cutoff is discontinuous at $R_\mathrm{cut}=R_\mathrm{d}$ 
with $\beta_\mathrm{s}^{(1/s)}\to 1$. We will use $s=4$ leading to $\beta_\mathrm{s}=0.70$ for
$\alpha=3/4$. In this case the surface density at $R_\mathrm{d}$ is $\Sigma(R_\mathrm{d})=0.49\,\Sigma_\mathrm{d}$.

An inspection of the scaling relations in NT shows that in the case of SMBHs
self-gravity of the AD for the vertical structure becomes important for
radii larger than $\sim 100 r_\mathrm{S}$. Since we cannot resolve the innermost
part of the AD we simplify the model of the vertical disk structure. We adopt a
self-gravitating isothermal profile given by
\bqn
\rho_\mathrm{g}(R,z)&=&\frac{\Sigma(R)}{\sqrt{2\pi}h_\mathrm{z}}
	\exp\left(-\frac{z^2}{2 h_\mathrm{z}^2}\right).
\label{eq-rhog1}
\eqn 
In the NT model the (half-)thickness $h_\mathrm{z}$ increases with distance according $\propto R^{9/8}$. 
Since this is altered by self-gravitation and since we cannot resolve the 
vertical structure of such a thin disc close to the inner boundary, 
we decided to adopt a constant thickness $h_\mathrm{z}$, 
taking an appropriate value at some intermediate radius of the AD.
To simplify the equations we define the dimensionless value $h =
h_z/R_\mathrm{d}$, which is also 
constant.

The effective sound speed $c_\mathrm{s}$ (which may be dominated by the turbulent motion in a clumpy gas) 
is given by (assuming a vertically isothermal model)
\bqn
c_\mathrm{s}^2(R)&=&8\pi G\rho_\mathrm{g}(R,0) h_\mathrm{z}^2=4\sqrt{2\pi}G\Sigma h_\mathrm{z}\label{eq-cs}\\
&=&\sqrt{\frac{8}{\pi}}(2-\alpha)\frac{\mu_\mathrm{d}}{\mu_\mathrm{bh}}
 h \left(\frac{R}{R_\mathrm{d}}\right)^{1-\alpha}
\,v_\mathrm{circ}^2(R),\nonumber
\eqn
where $v_\mathrm{circ}(R)$ is approximated by the Kepler rotation of the SMBH only. In a thin disk the radial pressure support can be neglected.
 Eq. \ref{eq-cs} shows that the rotation curve $v_\mathrm{circ}$ is highly supersonic ($v_\mathrm{circ}/c_\mathrm{s}\approx 100$ at the outer boundary $R_\mathrm{d}$ for the parameters used in our simulations: $\mu_\mathrm{bh}=0.1$, $\mu_\mathrm{d}=0.01$, $h=10^{-3}$). The Mach
number is increasing inwards.

The sound speed decreases with increasing radius and the stability of the AD is a function of radius. 
The Toomre $Q$ stability parameter is given by
\bq
Q^2=\left(\frac{\kappa c_\mathrm{s}}{\pi G \Sigma}\right)^2
=\frac{8}{2-\alpha}\sqrt{\frac{2}{\pi}}
\frac{\mu_{bh}}{\mu_{d}}h
\left(\frac{R}{R_\mathrm{d}}\right)^{\alpha-3}
\label{eq-Q}
\eq
with epicyclic frequency $\kappa$, which shows that heavy and thin ADs are unstable near the outer boundary. For the case of $\mu_{d}/\mu_{bh}=0.1$ and $h= 10^{-3}$ the AD is formally unstable at $R>0.26R_\mathrm{d}$. In the framework of our simplified AD model with constant thickness we may ignore this instability, since it could easily be avoided by adopting an increasing thickness with distance $R$. With eqs. \ref{eq-Sigma} and \ref{eq-Md} the density distribution of the AD with constant thickness (Eq. \ref{eq-rhog1}) is given by 
\bqn
\rho_\mathrm{g}(R,z)&=&\frac{2-\alpha}{2\pi\sqrt{2\pi}}\frac{M_\mathrm{d}}{h\, R_\mathrm{d}^3}
	\left(\frac{R}{R_\mathrm{d}}\right)^{-\alpha}\times \\
	&&\exp\left[-\beta_\mathrm{s}\left(\frac{R}{R_\mathrm{d}}\right)^s\right]
	\exp\left(-\frac{z^2}{2h^2R_\mathrm{d}^2}\right).\nonumber
\label{eq-rhog2}
\eqn 
%%%%%%%%%%%%%%%%%%%%%%%5
\begin{figure}[h]
\begin{center}
%\plotone{Fig1b}
\includegraphics[width=0.48\textwidth,angle=0]{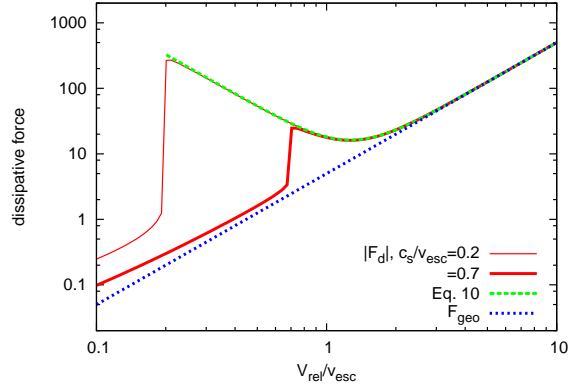}
\caption{Drag force $|F_\mathrm{d}|$ in units of
$[\pi{r_\mathrm{\star}^2}\rho_\mathrm{g}c_\mathrm{s}^2]$ for different sound
speeds $c_\mathrm{s}$ in the AD (full red lines). We have
chosen $Q_\mathrm{d}=5$ and $\ln\Lambda=13$. The dashed green line shows the
approximation given in Eq. \ref{eq-Fd} for the first case. $F_\mathrm{geo}$
(dotted blue line) is used in the simulations.} \label{figdyn}
\end{center}
\end{figure}
%%%%%%%%%%%%%%%%%%%%%%%5

\subsection{Dissipation of stellar kinetic energy} \label{sub-fric}

A detailed theory of the { dissipative} force working on the stars crossing
the AD { depends on the
details of} differential rotation, density profile,
turbulent motion in the disk and possible
resonance effects. { Stars interact with the AD typically many times
supersonically
before they are finally trapped in it. Hence,}  we restrict
our investigation to supersonic motion only, { neglecting the
details of the last few passages before trapping}. In this case we can
use for
the { dissipative} force the geometrical cross section
$F_\mathrm{geo}=Q_\mathrm{d}\pi{r_\mathrm{\star}^2}\rho_\mathrm{g}
V_\mathrm{rel}^2$  (determined by the effective area
$Q_\mathrm{d}\pi{r_\mathrm{\star}^2}$
of the bow shock with stellar radius $r_\mathrm{\star}$ and $Q_\mathrm{d}\sim
5$) enhanced by dynamical friction, which is the gravitational pull by the
over-density in the Mach cone due to gravitational focusing \citep{Ostriker:99}.
The { dissipative} force $\vec F_\mathrm{d}$ can be written in the form
\citep{Spurzem:04}
\begin{eqnarray}
\vec F_\mathrm{d}&=&-\pi{r_\mathrm{\star}^2}\rho_\mathrm{g}|\vec V_\mathrm{rel}|\vec V_\mathrm{rel}
\left[Q_\mathrm{d} +\left(\frac{v_\mathrm{esc}}{V_\mathrm{rel}}\right)^4\ln\Lambda\right]
	\nonumber\\&&\qquad\mbox{for }\quad V_\mathrm{rel}>c_\mathrm{s}.  
     \label{eq-Fd}
\end{eqnarray}
Here $v_\mathrm{esc}$ is the escape velocity at the
surface of the star ($v_\mathrm{esc}$=620\,km/s for the Sun) and
the relative velocity $\vec V_\mathrm{rel}=\vec V_\mathrm{\star}-\vec V_\mathrm{d}$ is the
velocity of the star in the frame co-rotating with the disc. 
$\Lambda$ corresponds to the length of the Mach cone in units of the
star radius $r_\mathrm{\star}$ leading to a Coulomb logarithm
$\ln\Lambda\sim 10-20$.

Fig.~\ref{figdyn} shows the { dissipative} force$|F_\mathrm{d}|$  for a range of
values of
$c_\mathrm{s}/v_\mathrm{esc}$ (full red lines). The dashed green line shows the
approximation given in Eq. \ref{eq-Fd} for $c_\mathrm{s}/v_\mathrm{esc}=0.2$.
The contribution to the dissipative force by $F_\mathrm{geo}$
is shown by the dotted blue line. From the strong dominance of the
dynamical friction for velocities $V_\mathrm{rel}$ smaller than $v_\mathrm{esc}$
we expect that stars are quickly decelerated to $V_\mathrm{rel}<c_\mathrm{s}$
and onto co-rotating circular orbits which then move slowly to the center with a
radial decay speed comparable to $c_\mathrm{s}$.
The { dissipative} force is anti-parallel to the relative velocity and can also be
accelerating with respect to the rest frame of the CSC.

%\underbar{ THREE PARAGRAPHS and THREE EQUATIONS taken out}

For a measure of the efficiency of the { dissipative} force we first introduce a
dissipative time scale $t_\mathrm{diss\star}$, which describes the energy
change ${\dot E}_\star^\mathrm{(sd)}$ { of a single star}
due to the { dissipative} force $F_\mathrm{d}$ arising from
interaction with the AD at the outer part of the disk. 
{ Our ansatz is:
\bq 
t_\mathrm{diss\star}
=\frac{\xi_\mathrm{k}  m_\star \sigma_\star^2 }{P_\mathrm{d} {\dot
E}_\star^\mathrm{(sd)}}
\label{eq-tdiss}
\eq 
with the stellar mass $m_\star$ and the 3D stellar velocity dispersion
$\sigma_\star$; the dissipative time scale as defined above depends strongly on
radius. At the outer edge $R_\mathrm{d}$ of the AD we have for example
${\dot E}_\star^\mathrm{(sd)} = Q_\mathrm{d} \pi r_\star^2 \rho_g \sigma_\star^3 $,
$\rho_g = \Sigma_\mathrm{d} / R_\mathrm{d} $,
%$\Sigma_\mathrm{d} = (2-\alpha) M_\mathrm{d} / (2 \pi R_\mathrm{d}^2)$, {\bf removed: not used here}
$t_\mathrm{dyn} = R_\mathrm{d} / \sigma_\star $,
and one gets
\bq
t_\mathrm{diss\star}(R_\mathrm{d}) =   \frac{\xi_k}{Q_\mathrm{d}P_\mathrm{d}}
\cdot \frac{\Sigma_\star}{\Sigma_\mathrm{d}} \cdot t_\mathrm{dyn}
\label{eq-tdiss2}
\eq
In the second form of the above equation we have defined the surface density
of stars $\Sigma_\star = m_\star / (\pi r_\star^2)$, 
which provides useful insights.
Note that this dissipative time scale related to a single star is a strongly increasing function with
radius, so the longest dissipation time can be found at the outer edge of the disk,
in our case at $R_d$.

We consider now the quantities $\xi_k$ and $P_d$. 
The former implicitly is defined
by $ 2 E_\star := \xi_k m_\star \sigma_\star^2(R_\mathrm{d}) = G m_\star M_\mathrm{cl} / R_\mathrm{d}$.
The latter, $P_\mathrm{d}$, accounts for the number of disk passages that
a star will need before its full kinetic energy is dissipated, and it
also includes an
average over the orbital parameters of the stars, which gives:}
\bq
P_\mathrm{d}=\left\langle g(e,i,R_\mathrm{d}/p)
        \frac{t_\mathrm{dyn}}{t_\mathrm{orb}}\right\rangle
\eq
The average in the equation above is taken over the disk crossing events of all stars 
with the proper efficiency function $g(e,i,R/p)$ which depends on the orbital
eccentricity $e$, the inclination with respect to
the AD $i$, and the focal parameter $p$  (the properties of $g(e,i,R/p)$
will be discussed in detail in a follow-up paper).
For simplicity we neglect here the contribution by
dynamical friction. It can easily be included by replacing $Q_\mathrm{d}$ with the velocity
dependent factor $\left(Q_\mathrm{d}+(v_\mathrm{esc}/V_\mathrm{rel})^4\ln\Lambda\right)$
in the definition of $P_\mathrm{d}$. 

%{\bf reordered and reformulated} For shorthand notation and discussion we define here the dimensionless time scale ratios
%$\delta = t_\mathrm{diss\star}/t_\mathrm{dyn}$ and $\eta_\star =
%t_\mathrm{diss\star}/t_\mathrm{rx}$ using
%Eq.~\ref{eq-tdiss2} and 
%{\bf removed}
%We get
%\bq
%\delta = \frac{\xi_\mathrm{k}}{Q_\mathrm{d} P_\mathrm{d}} \cdot
%\frac{\Sigma_\star}{\Sigma_\mathrm{d} } \ ; \qquad
%\eta_\star=\delta \cdot  \frac{\ln(0.4N)}{0.14 N}\ .
%\label{eq-eta}
%\eq
In energy space the dissipation process leads to a stationary flow of stars to
highly negative values; at some point stars will be
absorbed by the central black hole, for example by tidal disruption at the tidal radius $r_\mathrm{t}$. 
As discussed in \citet{VilkoviskiC:02} the dissipation process will bring stars down into
the plane of the disk, but two-body relaxation will reheat them in the vertical direction.
Both effects drive the
stars inwards to the black hole, where they are finally tidally disrupted and accreted.
Here we use a simple parametrized model; stars are absorbed onto
the central black hole at some accretion radius $r_\mathrm{acc}$, for which we
demand $R_\mathrm{d} \gg r_\mathrm{acc} \gg r_\mathrm{t}$, but otherwise
treat it as a free parameter, varied in our simulations later on. 

For scaling purposes with particle number $N$ in the simulations we use the standard half-mass relaxation time of \citet{Spitzer:87}:
\begin{eqnarray}
t_\mathrm{rx}&=&\frac{0.14N}{\ln(0.4 N)}t_\mathrm{dyn} \quad \mbox{with} \quad t_\mathrm{dyn}=\sqrt{\frac{r_\mathrm{half}^3}{GM_\mathrm{cl}}}
\label{eq-trx}
\end{eqnarray}
where $r_\mathrm{half}$ is the half-mass radius of the CSC.
The half-mass relaxation time $t_\mathrm{rx}$ is given in
Table \ref{tab-scaling} for a series of galactic
nuclei covering the observed range of SMBH masses.

The local relaxation time 
$t_\mathrm{rc}$ due to two-body encounters is given by \citep[][Eq. (8-71)]{BT87}
\begin{eqnarray}
t_\mathrm{rc}&=&\frac{0.34\sigma^3}{G^2m_\mathrm{s}\rho_\mathrm{s}\ln\Lambda}\\ \nonumber
&=&\frac{18\mathrm{Gyr}}{\ln\Lambda}
\left(\frac{\sigma}{1\mathrm{km\,s^{-1}}}\right)^3
\left(\frac{1\Msol}{m_\mathrm{s}}\right)
\left(\frac{1\Msol\mathrm{pc}^{-3}}{\rho_\mathrm{s}}\right)     \label{eq-trc}
\end{eqnarray}
with 1-D velocity dispersion $\sigma$, mean particle mass $m_\mathrm{s}$, and
density $\rho_\mathrm{s}$ of the CSC. This equation can be applied at each
radius or to the central region of a system by using mean values inside some
radius $r$ and is also called the 'core relaxation time'. 
Inside the gravitational influence radius $r_\mathrm{h}$ %{\bf variabel included for definition}
of the
black hole (which is in our model of the same order as
the disk outer radius $R_d$) we assume $\sigma_\star^2 \propto r^{-1}$ and the
standard Bahcall-Wolf
density cusp solution for the stellar density profile in the CSC is $\rho\propto r^{-7/4}$
\citep{BahcallW:76}.
Hence we get for the local relaxation time (cf. Eq.~\ref{eq-trc})
$t_\mathrm{rc}\propto r^{1/4}$. %{\bf corrected: t_rx to t_rc}
For the galactic nuclei in Table \ref{tab-scaling} the core relaxation time
at the influence radius $t_\mathrm{rc}(r_\mathrm{h})$ is a factor of 1.5 to 2 shorter than $t_\mathrm{rx}$. This shows that 2-body relaxation in the central part of galactic nuclei is well represented by our confined model with a CSC mass only 10 times larger than the SMBH mass.

For comparison with measurements in our simulations, as described later,
it is more practical to define a {\em global} dissipation time scale $t_\mathrm{diss}$ and
a {\em global} dimensionless dissipative time scale $\eta$ in the following way:
\bqn
\eta &=& t_\mathrm{diss}/t_\mathrm{rx} \\ \nonumber
t_\mathrm{diss} &=& \frac{E_k}{\dot{E}^{(sd)}} =
\frac{M_\mathrm{cl} \sigma_\star^2(r_\mathrm{acc})}{2\dot{E}^{(sd)}}
=\frac{G M_\mathrm{bh} M_\mathrm{cl}}
{2r_\mathrm{acc}\dot{E}^{(sd)}}
\label{eq-eta2}
\eqn
Here $E_k$, with $2 E_k = M_\mathrm{cl} \sigma_\star^2(r_\mathrm{acc}) = G
M_\mathrm{cl}M_\mathrm{bh} / r_\mathrm{acc}$, 
and $\dot{E}^{(sd)}$ are the total kinetic energy and total energy
dissipation
rate of all stars at the accretion radius. 
An accretion time scale for the black hole growth is defined by $t_\mathrm{acc} = 
M_\mathrm{bh}/{\dot M}_\mathrm{bh}$, or in dimensionless form $\nu = t_\mathrm{acc}/t_\mathrm{rx}$.
We remove stars at $r_\mathrm{acc}$; so in a stationary state 
the energy dissipation
rate required at $r_\mathrm{acc}$ to sustain the black hole mass accretion rate is
${\dot E}^\mathrm{(sd)} = {\dot M}_\mathrm{bh} GM_\mathrm{bh}/(2 r_\mathrm{acc})$.
With equation \ref{eq-eta2}
we get
\bq
\nu=\frac{t_\mathrm{acc}}{t_\mathrm{rx}}=\eta \frac{t_\mathrm{acc}}{t_\mathrm{diss}}
 = \eta \quad \frac{M_\mathrm{bh}}{{\dot M}_\mathrm{bh}} \quad \frac{{\dot E}^\mathrm{(sd)}}{E_k}
 = \eta \mu_\mathrm{bh}
\label{eq-nu}
\eq
Hence the black hole mass accretion rate does not depend on the choice of
$r_\mathrm{acc}$, as will be verified in our numerical simulations later.

%----------------------------------------------------------------------------%
\begin{table*}
\caption{Some examples for the physical scaling of galactic nuclei.
\label{tab-scaling}}{\footnotesize
\hspace*{-0.4cm}\begin{tabular}{*{10}{c}}
\tableline
Object	&$M_\mathrm{bh}$ [$\Msol$] 	& $\sigma_\mathrm{c}$ [km/s] & $r_\mathrm{h}$[pc]	&$N$	&$t_\mathrm{dyn}$ [Myr] 	& $t_\mathrm{rx}$[Gyr]  & $Q_\mathrm{tot}$	& $Q_\mathrm{tot}(8k)$ &$L_\mathrm{max}$ [$\Lsol$]\\
\tableline
M~87	&$6.6\times 10^9$&312	&291&$6.6\times 10^{10}$&1.5	&$6\times 10^{5}$	&$2.1\times 10^{-9}$	      &$5.7\times 10^{-3}$      &$1\times 10^{8}$ \\
NGC 3115&$9.6\times 10^8$&230	&78	&$9.6\times 10^9$&0.58	&$3.4\times 10^{4}$	&$4.2\times 10^{-9}$	      &$1.8\times 10^{-3}$      &$2.8\times 10^{7}$ \\
NGC 4291&$3.2\times 10^8$&242	&24	&$3.2\times 10^9$&0.16	&3400	&$1.5\times 10^{-8}$	      &$2.4\times 10^{-3}$      &$9.4\times 10^{7}$ \\
M 31	&$1.5\times 10^8$&160	&25	&$1.5\times 10^9$&0.26	&2690	&$6.2\times 10^{-9}$	      &$4.7\times 10^{-4}$      &$5\times 10^{7}$ \\
NGC 4486A&$1.3\times 10^7$&111	&4.5	&$1.3\times 10^8$&0.067	&68.8	&$1.7\times 10^{-8}$	      &$1.2\times 10^{-4}$      &$1.9\times 10^{8}$ \\
MW	&$4\times 10^6$	&110	&1.4	&$4\times 10^7$	&0.021	&7.2	&$5.2\times 10^{-8}$	      &$1.2\times 10^{-4}$      &$5\times 10^{8}$ \\
M 32	&$3\times 10^6$	&75	&2.3	&$3\times 10^7$	&0.050	&12.9	&$1.5\times 10^{-8}$	      &$2.8\times 10^{-5}$      &$2\times 10^{8}$ \\

\end{tabular}
}
\tablecomments{ 
For the CSC we adopt $\mu_\mathrm{bh}=0.1$, solar-type
stars with $m_\mathrm{\star}=1\Msol$, $r_\mathrm{\star}=2.3\times  10^{-8}$\,pc, 
$v_\mathrm{esc}=620$\,km/s,
$R_\mathrm{d}=r_\mathrm{h}$ and $r_\mathrm{half}=3\,r_\mathrm{h}$, and $Q_\mathrm{d}=5$ for the bow shock size. 
Most observational data (Cols. 2 and 3) are taken from \citet{Gueltekin:09}, improved values for M~87 are from \citet{Murphy:11}, and $M_\mathrm{bh}$ for the Milky Way (MW) is from \citet{Gillessen:09}.
The influence radius (Col. 4) is derived from $r_h = GM_\mathrm{bh}/\sigma_c^2$, with
the core velocity dispersion $\sigma_c$ of the CSC; $N$ is the number of stars in the CSC derived
from $M_\mathrm{bh}/\mu_\mathrm{bh}$, 
the dynamical and half-mass relaxation timescales (Cols. 6 and 7) are derived by Eq. \ref{eq-trx}, the physical and scaled cross sections $Q_\mathrm{tot}$ and $Q_\mathrm{tot}(8k)$ (Cols. 8 and 9) by Eqs. \ref{eq-Qtot} and \ref{eq-Qscale},
and the maximum luminosity by accretion (Col. 10) assuming $L_\mathrm{max}=0.1\dot{M}_\mathrm{bh}c^2$.}
\end{table*}
%----------------------------------------------------------------------------%

\section{The Numerical Model} \label{sec-model}

%This is the first time that such high precision direct N-body 
%simulation considers non-gravitational forces. However, our 
%disk is yet stationary, with a Keplerian rotation velocity. 
%We include as free parameters the cross section of the ram 
%pressure force and the accretion radius at the inner edge 
%of the disk.

%We have accomplished the inclusion of a standard ram pressure force 
%into the Hermite scheme of the N-body code $\phi$GRAPE. 

For the high resolution direct $N$-body simulations of the CSC we used the 
specially developed { $\phi$GRAPE} (= { P}{\it arallel} 
{ H}{\it ermite} { I}{\it ntegration} with { GRAPE}) code. 
The program uses the 4-th order Hermite integration scheme for 
the particles with hierarchical individual block time-steps, 
together with the parallel usage of GRAPE6 (or nowadays the { GPU} - 
{ G}raphics { P}rocessing { U}nit) cards for the hardware calculation of 
the acceleration ${\vec  a}$ and the first time derivative 
${\dot{{\vec a}}}$ of the acceleration. The code itself and also 
the special GRAPE hardware is described in more detail in \citet{Hetal2007}. 
For all the new calculations we use the different 
NVIDIA CUDA/GPU hardware which emulate the standard GRAPE 
library calls\footnote{\tt ftp://ftp.mao.kiev.ua/pub/users/berczik/STARDISK/}.
%with the {\tt SAPPORO} \citet{sapporo} or {\tt YEBISU} \citet{yebisu} 

Here we mention briefly the most important features 
added to the Hermite scheme of the N-body $\phi$GRAPE code
in order to model the ram pressure { dissipative} force 
and the star accretion to the SMBH:

\begin{itemize}
\item A { dissipative} force routine, where we calculate the acceleration caused by the 
interaction with the gaseous disk $\vec a_\mathrm{d}$ (Eq. \ref{eq-Fd2}) 
and its first time derivative ${\dot{{\vec a}}}_\mathrm{d}$. 
\item We reduce the time-steps of stars when they come close to the disk plane. 
 { Otherwise} stars with big individual time step may miss the disk 
and would not feel the effect of the 
disk at all.
\item A simple accretion scheme of stars onto the SMBH, { where the stellar
mass is added to the central black hole if they get inside a certain
accretion radius. The accretion radius is used as a free scaling parameter;
results for different accretion radii can be used to scale to real parameters
of galactic nuclei (see \ref{sec-accretion}).}
This algorithm has been
described and used in \citet{Jose:2011, Li:2012}. 
\item In order to control integration error we add a careful bookkeeping of 
energy changes caused by removing stars in process of accretion 
and by the { dissipative} force of the stars-gas interaction. 
In all our runs the total energy error does not exceed $10^{-4}$.
\end{itemize}

In the simulations we use standard $N$-body units given by
\bq
G=M_\mathrm{cl}=4|E_\mathrm{tot}|=1, 
\eq
where $E_\mathrm{tot}$ is the total energy of the initial Plummer sphere for
the CSC without a perturbing SMBH
\citep{HeggieM:86}. { Note that in this scaling an increase of the particle number takes
place at constant total mass; hence the
stellar mass $m_\star \propto 1/N$ decreases with particle number, and it also decreases with 
respect to e.g. the disk mass and black hole mass.}

Our model consists of three components: the CSC, the SMBH, and the AD. The CSC with initial 
mass $M_\mathrm{cl}$ is realized by $N$ particles of mass $m_\star$.
{ The initial density distribution is} 
 a Plummer model, { which is only} in dynamical equilibrium 
{ if the gravity} of the central SMBH { is ignored}. 
{ The system adjusts quickly in a few dynamical time scales.}
The mutual gravitational interaction{s} and the { dissipative} force { on the stars}
in the AD are calculated { fully}
by the $\phi$GRAPE code.

The SMBH with initial mass $M_\mathrm{bh} = \mu_\mathrm{bh}M_\mathrm{cl}$ is modeled by an analytic Kepler potential fixed at the origin. We allow accretion of stars, which are effectively captured by the
inner part of the AD or scattered by two-body relaxation into the loss cone, to grow the mass of the SMBH.
Physically the accretion radius $r_\mathrm{acc}$ is given by the tidal radius $r_\mathrm{t}$, where the stars are disrupted, or the Schwarzschild radius $r_\mathrm{S}$. Both radii are orders of magnitude below our numerical resolution. Therefore we need to analyze the scaling of the accretion with decreasing accretion radius. The orbits of the stars accreted by the interaction with the AD are circularized before being accreted. In contrast, stars accreted by two-body scattering into the loss cone are predominantly accreted on hyperbolic orbits. In order to simulate the effect of the different eccentricity distribution, we apply a second accretion criterion based on the velocity of the stars. We define an accretion criterion to merge stars with the SMBH using two criteria:  1) distance $r<r_\mathrm{acc}$ and 2)  $V_\mathrm{\star}^2<k_\mathrm{acc}v_\mathrm{circ}^2$. Stars well inside the influence radius of the SMBH are moving on (local) Kepler orbits, where the velocity is given by $V_\mathrm{\star}^2=v_\mathrm{circ}^2(2-r/a)$ with semi-major axis $a$.
In the limit of large $k_\mathrm{acc}$ all stars reaching $r_\mathrm{acc}$ would be accreted. For $k_\mathrm{acc}=1$ stars inside $r_\mathrm{acc}$ are accreted, if  $a\le r_\mathrm{acc}$, i.e. all stars with energy below $-GM_\mathrm{cl}/r_\mathrm{acc}$ (neglecting the potential energy of the CSC) are accreted in one orbital time. In all simulations we use $k_\mathrm{acc}=1$. 
%Due to technical reasons with the time resolution of encounters close to the SMBH on hyperbolic orbits
%combined with strong friction forces, we accrete additionally all stars if $r<r_\mathrm{acc}/10$ independent
%of the semi-major axis for the simulations with drag force. We have tested that the accretion rates are not
%changed significantly by this additional accretion criterion. 
(Note for completeness that in some runs with AD we accreted all stars with $r<r_\mathrm{acc}/10$ independent of their energy.  We have tested that the accretion rates are not changed significantly by this additional accretion criterion.)

%Since the friction in the inner part of the disk is very effective 
%to move the stars to the SMBH, the artificially large $r_\mathrm{acc}$ 
%leads to a small time delay in the accretion events only. 

The properties of the AD are fixed by the normalized mass $\mu_\mathrm{d}$ with analytic density distribution
according to Eq. \ref{eq-rhog1} with $\alpha=3/4$, $s=4$, and constant thickness $h=10^{-3}$. We use a Kepler rotation of the AD in the potential of the SMBH neglecting the gravity of the CSC and pressure gradients in the AD. We set the outer cutoff initially at $R_\mathrm{d}= r_\mathrm{h}$. 
Using a Kepler rotation curve underestimates the real circular
speed at the outer boundary by a factor of 1.4, but this has no
significant influence on the dynamics of the stars. The reason is that the
gas density and thus the friction force is very small in the outer regions
of the AD. The part of the AD with significant dissipation of energy of
crossing stars is deep inside the influence radius of the SMBH. Here the
approximation of Kepler rotation is very good. We have chosen the large
cutoff radius only in order to avoid another free parameter.

It is helpful { for calibration of our models with respect to real systems in
galactic nuclei} to define an effective dissipative parameter
\bq
Q_\mathrm{tot}= Q_\mathrm{d} \frac{N\pi r_\mathrm{\star}^2}{\pi R_\mathrm{d}^2}.
\label{eq-Qtot}
\eq
Note that $Q_\mathrm{tot}$ enhances $ Q_\mathrm{d}$ by a factor,
which describes the dimensionless total dissipative
cross section of $N$ stars, normalized to the disk area.
With this definition we can rewrite the original dissipative force
in Eq. \ref{eq-Fd} as acceleration
\bq
\vec a_\mathrm{d} = \vec F_\mathrm{d} /m_\star =-Q_\mathrm{tot}\frac{\pi R_\mathrm{d}^2\rho_\mathrm{g}}{M_\mathrm{cl}}|\vec V_\mathrm{rel}|\vec V_\mathrm{rel}.
\label{eq-Fd2}
\eq
{ The {\em local} parameters like $r_\star$, $m_\star$, and $Q_d$, whose scaling in
terms of $N$-body units is not straightforward, are transformed in this
way to the {\em global} quantities such as $R_\mathrm{d}$, $M_\mathrm{cl}$, and
$Q_\mathrm{tot}$.

Due to numerical limitations, direct $N$-body simulations are performed here
with particle numbers smaller than that in a real galactic nucleus. This
leads to the relaxation time being a function of the particle number, $N$,
whereas it is a fixed quantity for a given nucleus. To correct for this,
the simulations are run for a multiple of the relaxation time and are then
rescaled to the correct time units to compare to a physical galactic
nucleus. for this reason, all physical processes, including the star-disc
drag, need to be rescaled as well.
In our models all quantities except $Q_\mathrm{tot}$ are invariant
when changing the particle
number in the simulation.
Therefore for calibration to real systems we just have to compute the value
of $Q_\mathrm{tot}$ and adjust it correspondingly in our model system.
But this step alone is not sufficient; our model system has a much shorter
two-body relaxation time than in reality. We want to study numerical systems
at smaller particle number, which have all the same ratio $\eta$ of dissipation to
relaxation time scale.} 
Therefore we need to choose $Q_\mathrm{tot}$ 
such that the same value for $\eta$ is achieved: 
\bq
Q_\mathrm{tot}(N_2)=\frac{t_\mathrm{rx}(N_1)}{t_\mathrm{rx}(N_2)}Q_\mathrm{tot}(N_1)
\label{eq-Qscale}
\eq
Our model aims to discover numerically the secular evolution of the
 coupled star gas system in galactic nuclei. Therefore we are interested
 in the interplay between dissipative processes (star-gas drag) and
 two-body relaxation processes - the semi-analytic model of \citet{VilkoviskiC:02} defines the setup in which we want to put our numerical
 models. Since we are not yet able to simulate star-by-star, two different
 effects have to be taken into account for this: (i) each simulation
particle
 effectively represents many stars, not a single star. Therefore we cannot
 study the individual star-gas drag effects here, only the collective one;
 this is shown by the definition of $Q_\mathrm{tot}$ in Eq. \ref{eq-Qtot}, which shows how for
one
 simulation particle an effective cross section is realized, which
corresponds
 to the effect of the much larger stellar system we have in mind. (ii)
 In our system with lower particle number the two-body relaxation time has
 a ``wrong'' ratio to other time scales. Therefore, we rescale $Q_\mathrm{tot}$ in
 such a way, that the relaxation time and the dissipation time have the
 correct ratio for the larger system, which is done in the Eq. \ref{eq-Qscale}.
{ Eq. \ref{eq-Qscale} shows, for example, that} for a system 
with $10^8$ particles, to be simulated by only $N=10^4$,
we need to increase $Q_\mathrm{tot}$ 
artificially by a factor of $\sim 5,000$. In Table~\ref{tab-scaling} 
the real values of $Q_\mathrm{tot}$ and simulated values for $N=8000$ are both given 
for the selected galaxies.
 It should be stressed that we do NOT
intend
 to model star-gas interactions in detail. Rather we apply a standard
model
 of it, scale it up in a collective way and calibrate it with respect to
 the relaxation time. Our main goal is to check the increase of star
accretion rates due to the presence of a central AD.

In our starting configuration we added the central SMBH to the Plummer distribution of the CSC. In order to start the analysis of the system in dynamical equilibrium we let the CSC evolve for $\sim 5 t_\mathrm{dyn}$ with the SMBH. After the initializing phase the cusp around the SMBH shows a density slope $\gamma\approx 2.5$ which still differs from a BW cusp (top panel of Fig.~\ref{figMcum}). The difference of the set-up Plummer profile and the cumulative mass profile at $t=0$ demonstrates the effect of virialization in the potential of the SMBH. Then we determine $R_\mathrm{d}$ such that at $t=0$ the enclosed CSC mass at $R=R_\mathrm{d}$ equals the SMBH mass.

%----------------------------------------------------------------------------%
\begin{table*}
\caption{Model parameters of all runs.
\label{tab-model-list}}
\begin{tabular}{r*{7}{c}}
\tableline
 Model &  $N$ & $\epsilon/R_\mathrm{d}$  & $r_\mathrm{acc}/R_\mathrm{d}$ & $Q_\mathrm{tot}(N)$\\
\tableline
 04K-005 		& $4k$	&$4.55 \times 10^{-4}$			&$5\times 10^{-3}$	&$10^{-2}$			& \\
 04K-01			& $4k$	&$4.55 \times 10^{-4}$			&$10^{-2}$		&$10^{-2}$			& \\
 04K-02			& $4k$	&$4.55 \times 10^{-4}$			&$2\times 10^{-2}$	&$10^{-2}$			& \\
 04K-04			& $4k$	&$4.55 \times 10^{-4}$			&$4\times 10^{-2}$	&$10^{-2}$			& \\
 08K-005 		& $8k$	&$4.55 \times 10^{-4}$			&$5\times 10^{-3}$	&$5.47\times 10^{-3}$	& \\
 08K-01			& $8k$	&$4.55 \times 10^{-4}$			&$10^{-2}$		&$5.47\times 10^{-3}$	& \\
 08K-02			& $8k$	&$4.55 \times 10^{-4}$			&$2\times 10^{-2}$	&$5.47\times 10^{-3}$	& \\
 08K-04			& $8k$	&$4.55 \times 10^{-4}$			&$4\times 10^{-2}$	&$5.47\times 10^{-3}$	& \\
 16K-005		& $16k$	&$4.55 \times 10^{-4}$			&$5\times 10^{-3}$	&$2.97\times 10^{-3}$	& \\
 16K-01			& $16k$ &$4.55 \times 10^{-4}$			&$10^{-2}$		&$2.97\times 10^{-3}$	& \\
 16K-02			& $16k$ &$4.55 \times 10^{-4}$			&$2\times 10^{-2}$	&$2.97\times 10^{-3}$	& \\
 16K-04			& $16k$ &$4.55 \times 10^{-4}$			&$4\times 10^{-2}$	&$2.97\times 10^{-3}$	& \\
\end{tabular}
\tablecomments{ 
Column 1 gives the identification label of the model, 
columns 2--4 are number of particles, smoothing length 
and accretion radius. Column 5 gives the total cross section, $Q_\mathrm{tot}$,
scaled according to Eq. \ref{eq-Qscale}.
Common parameters for all models are $\mu_\mathrm{bh}=0.1$,
$\mu_\mathrm{d}=0.01$, $h=10^{-3}$, $k_\mathrm{acc}=1$. }
\end{table*}
%----------------------------------------------------------------------------%

We performed a series of simulations combining different particle numbers
$N=4k,\,8k,\,16k$ and different accretion radii
$r_\mathrm{acc}=0.04,\,0.02,\,0.01,\,0.005$ (see Table~\ref{tab-model-list})
until  $t=10 t_\mathrm{rx}$.
The identifiers of the simulations in Table~\ref{tab-model-list} are coding the
particle number and accretion radius. For each 
parameter combination a run without { dissipative} force is done for comparison.
For the different $N$ 
we applied the scaling according to equation \ref{eq-Qscale}. For the 
simulations with { dissipative} force { our choice of} 
$Q_\mathrm{tot}$ corresponds to the case of M~87 ({compare }
Tables \ref{tab-scaling} and \ref{tab-model-list}). 
From the last column in Table \ref{tab-scaling} it is obvious that for lower
mass black holes the value of $Q_\mathrm{tot}$ in our simulations is
unrealistically large.

For the calculation of the CSC dynamics we used the parallel GRAPE systems 
built at the Astronomisches Rechen-Institut in 
Heidelberg\footnote{GRACE: {\tt http://www.ari.uni-heidelberg.de/grace}}, 
and at the Fesenkov Astrophysical Institute in Almaty. The code has recently been ported to large clusters with GPU hardware (in Beijing, China and Heidelberg, Germany, see acknowledgments) and results from these facilities have partly been used for this project. 

%The numerically most expensive simulations were performed on the 
%LAOHU GPU-cluster at the NAOC in Beijing ???.

\section{Results} \label{sec-result}

\begin{figure}
\includegraphics[width=0.48\textwidth,angle=0]{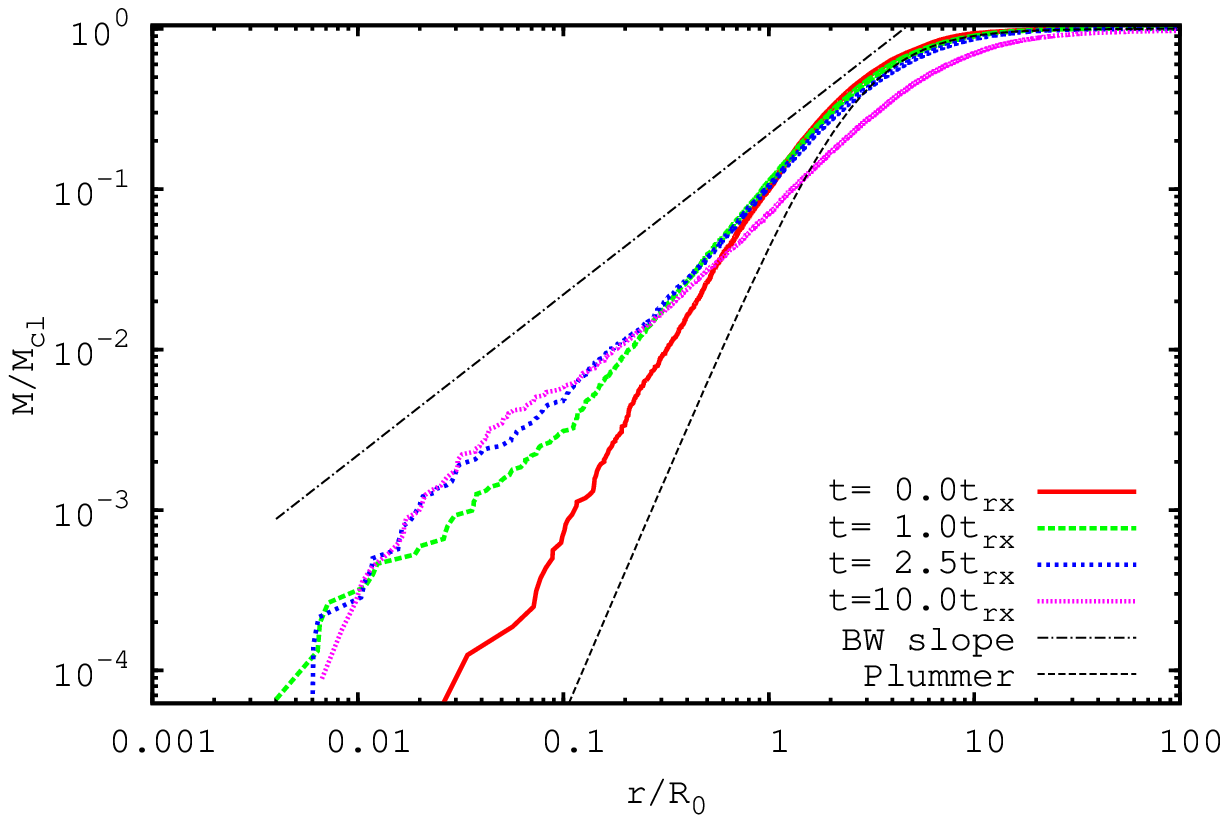}\\
\includegraphics[width=0.48\textwidth,angle=0]{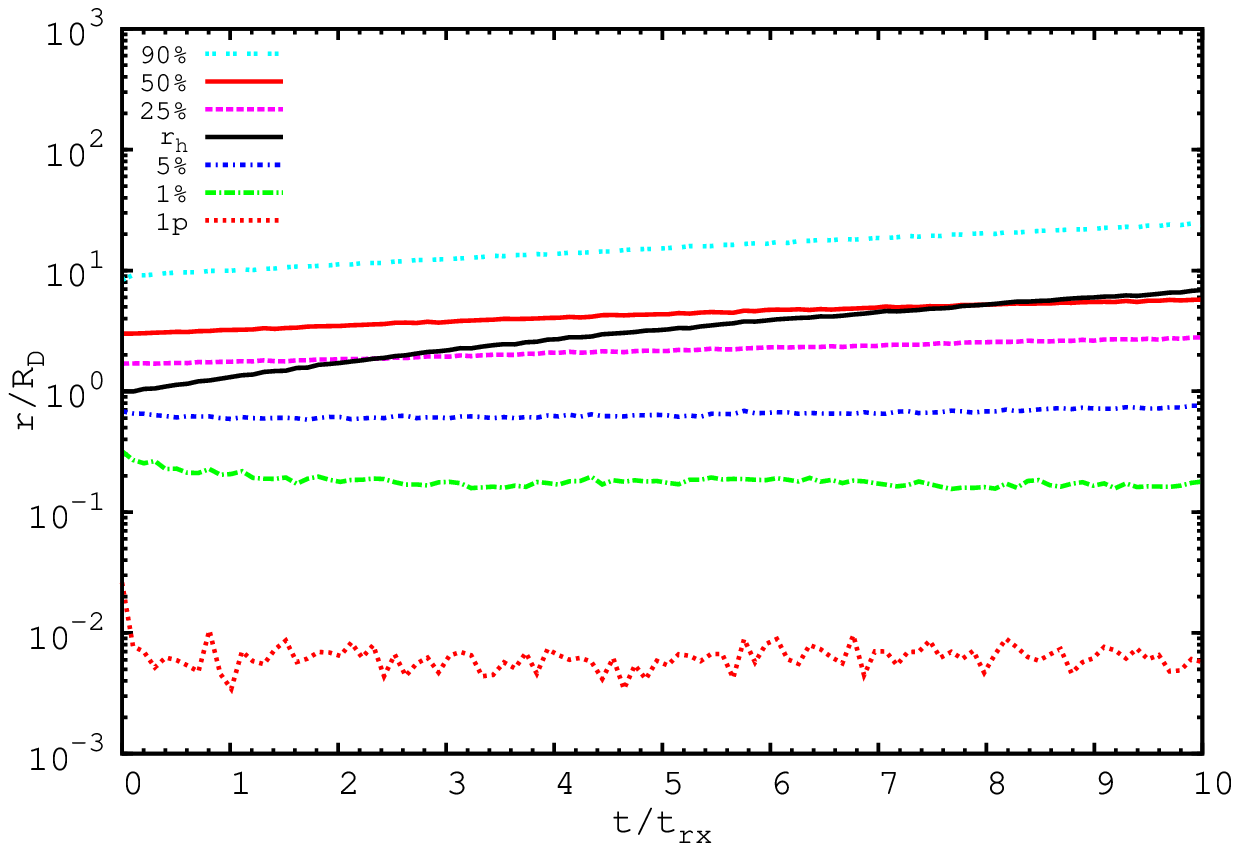}
\caption{Evolution of the CSC for model 16K-005. Top panel: Cumulative mass profiles in units  of the initial CSC mass at
different times with the Plummer profile and the power law of a
Bahcall-Wolf (BW) cusp added for comparison.
Bottom panel: Lagrange radii enclosing a fixed mass fraction with respect to
$M_\mathrm{cl}(t)$. The position of the inner-most particle (1p) and the influence radius
$r_\mathrm{h}$ are also shown.} 
\label{figMcum}
\end{figure}

In our starting configuration the CSC is in dynamical equilibrium including the
gravitational potential of the SMBH and has a steep cusp in the density profile.
The cumulative mass profiles in the top panel of Fig.~\ref{figMcum}) shows that
after one relaxation time the BW cusp is in place and remains stable over the
full simulation. The lower panel of Fig.~\ref{figMcum} shows the Lagrange radii
of the CSC for 1, 5, 25, 50, 90, 99 percent enclosed mass with respect to the
total mass $M_\mathrm{cl}(t)$ for the model 16K-005 (see
Table~\ref{tab-model-list} for model parameters). 
Additionally the position of the innermost particle (1p) shows that stars
crossing $r_\mathrm{acc}$ on highly eccentric orbits are quickly circularized
and accreted. The influence radius $r_\mathrm{h}$ of the SMBH is increasing
during the simulation by an order of magnitude due to accretion and the size of
the Bahcall-Wolf cusp is growing accordingly.
%%Fig.2%%%%%%%%%%%%%%%%%%%
%%%%%%%%%%%%%%%%%%%%%

\subsection{Scaling with particle number} \label{sec-timescales}

In simulations without AD, stars reaching $r_\mathrm{acc}$ are
immediately accreted onto the SMBH. The corresponding loss-cone in phase space
has a maximum angular momentum for each energy defining the opening angle of the
loss cone in phase space. The regime of an empty loss cone depends on
$t_\mathrm{rx}/t_\mathrm{dyn}$ and the width of the loss cone \citep{Amaro:01}.
In the regime with a high probability of a star to be scattered in or out of the
loss-cone, the loss cone is full.
In the empty loss-cone regime all stars scattered by two-body encounters into the loss cone are accreted and the accretion rate scales with $t_\mathrm{rx}$. In the full loss cone regime the accretion rate depends on $t_\mathrm{dyn}$. In a given system the empty part of the loss cone increases with increasing particle number $N$ leading to a $N$ dependence of the accretion rate. We used an additional energy criterion accreting stars with semi-major axis $a<r_\mathrm{acc}$, which cuts the loss-cone in energy space. 

\begin{figure}
%\plottwo{Fig3tp-mass-bh-04}{Fig3bt-mass-bh-005}
\includegraphics[width=0.48\textwidth,angle=0]{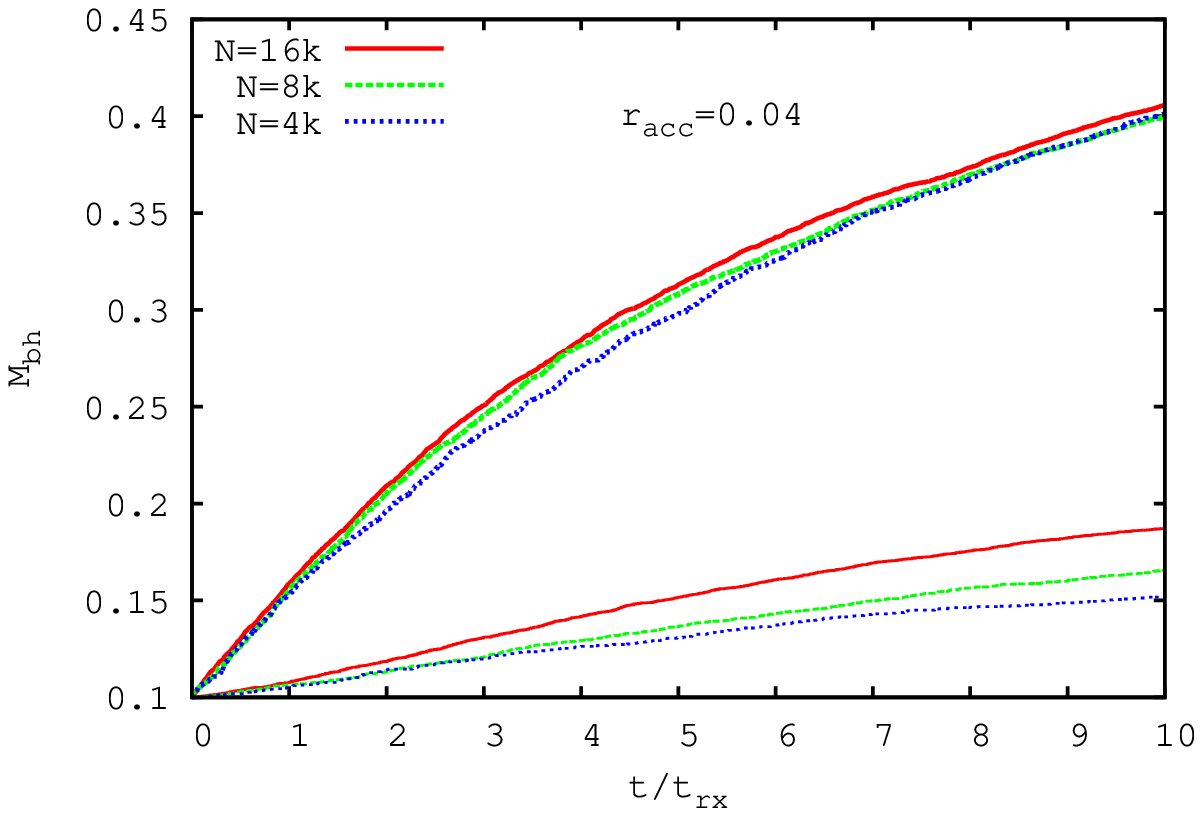}\\
\includegraphics[width=0.48\textwidth,angle=0]{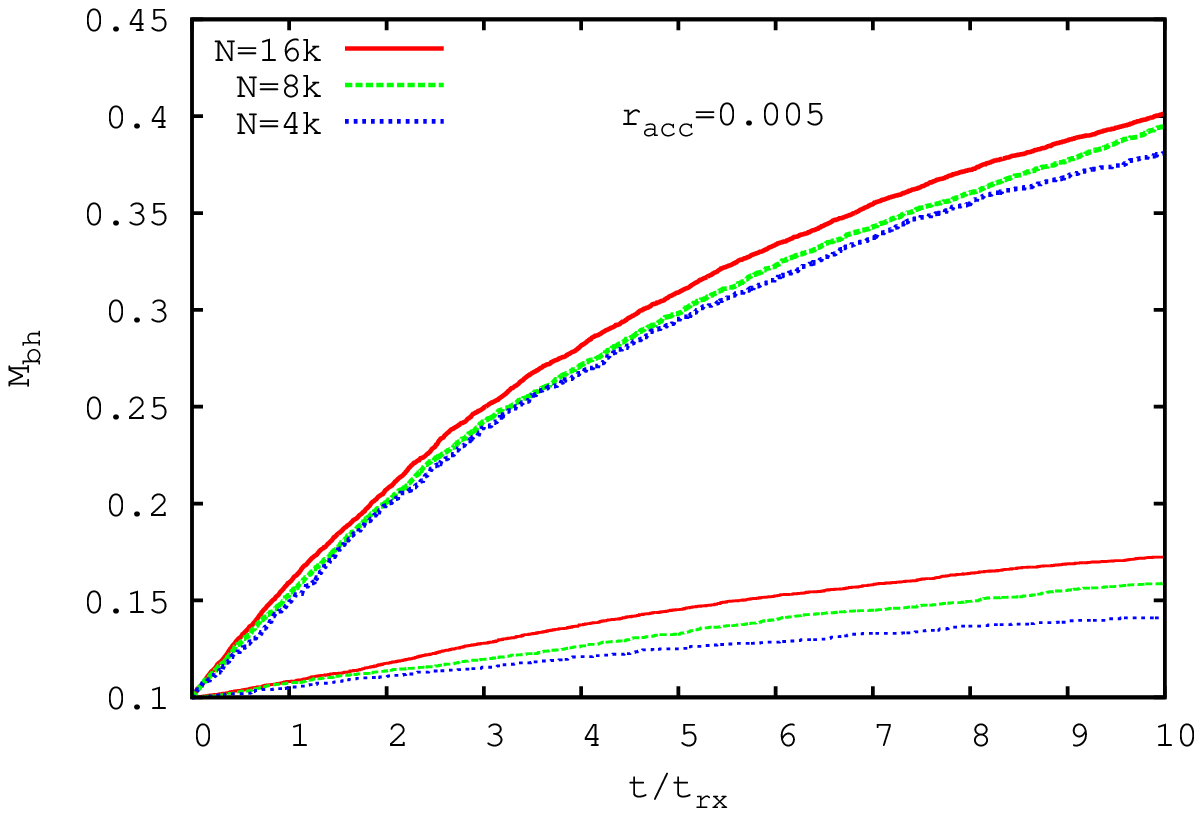}
\caption{SMBH growth with (upper thick lines) and without (lower thin lines) {dissipative} force for different accretion radii $r_\mathrm{acc}=0.04$ (top panel) and $r_\mathrm{acc}=0.005$ (bottom panel). The particle number ranges from $N$=4k...16k. 
$Q_\mathrm{tot}$ scales according to Eq. \ref{eq-Qscale} with $N$.} 
\label{figNoGas}
\end{figure}

In Fig.~\ref{figNoGas} the lower set of thin lines show the growing mass of the
SMBH for different particle numbers for the cases without dissipation due to the
AD (top panel: $r_\mathrm{acc}=0.04$ and bottom panel: $r_\mathrm{acc}=0.005$).
The accretion rate per relaxation time is seen to slowly increase with particle
number.
%%Fig.3%%%%%%%%%%%%%%%%%%%
%%%%%%%%%%%%%%%%%%%%%%%
The upper set of thick lines in Fig.~\ref{figNoGas} show that accretion is
significantly larger when
the { dissipative} force of the AD is included. The
accretion rate is found to be independent of particle number $N$.
%%Fig.4%%%%%%%%%%%%%%%%%%%
\begin{figure}
\includegraphics[width=0.48\textwidth,angle=0]{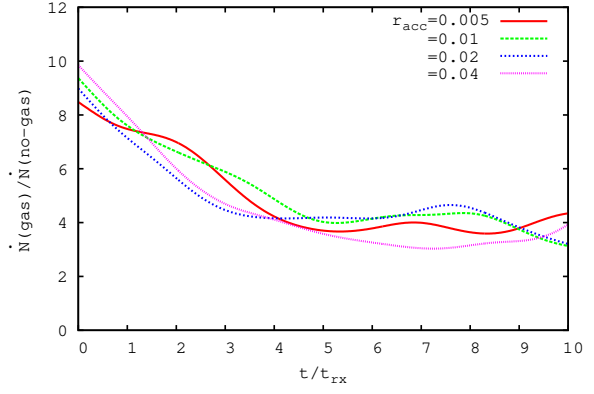}
\caption{Ratio of the accretion rates with and without {dissipative} force for $N$=8k and different
 $r_\mathrm{acc}$.} 
 \label{figNaccQ01overNaccQ00}
\end{figure}
%%%%%%%%%%%%%%%%%%%%%%%
In Fig.~\ref{figNaccQ01overNaccQ00} the ratio of the accretion rates with and
without { dissipative} force is quantified for the $8k$ simulations at different
accretion radii. After a few relaxation times an equilibrium with an enhancement
factor of $\sim 4$ is established for all accretion radii.

\subsection{The accretion radius} \label{sec-accretion}

\begin{figure}
\includegraphics[width=0.48\textwidth,angle=0]{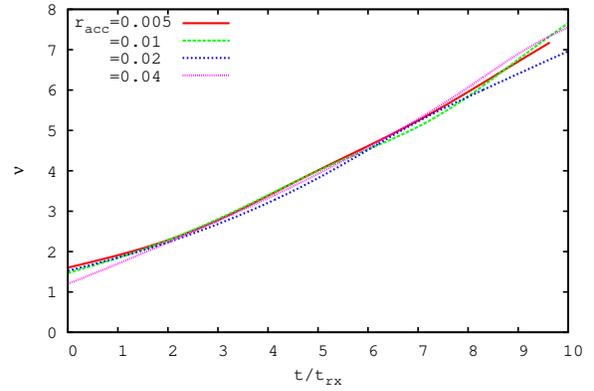}
\caption{Normalized accretion timescale $\nu$ (Eq. \ref{eq-nu}) for $N=16k$ with different $r_\mathrm{acc}$ (same line-styles as in Fig.~\ref{figNaccQ01overNaccQ00}).
}
\label{figNoverNdot}
\end{figure}

The physical accretion radius $r_\mathrm{acc}$ is much smaller than the numerical resolution.
Therefore the scaling of the accretion rate with $r_\mathrm{acc}$ is very important.
Without the AD the standard loss cone becomes thinner with decreasing $r_\mathrm{acc}$, leading to a decreasing accretion rate in the limit of an empty loss cone \citep[e.g.][]{Amaro:01}. On the other hand the Bahcall-Wolf cusp is characterized by a constant mass and energy flow to the SMBH and high binding energy, respectively. As a consequence the accretion rate based on an energy criterion should be independent of $r_\mathrm{acc}$. In Fig.~\ref{figNoverNdot} the independence of the accretion rate in terms of the accretion timescale $\nu$ (Eq. \ref{eq-nu}) on $r_\mathrm{acc}$ is shown for all $N=16k$ runs with { dissipative} force. For $t>1\,t_\mathrm{rx}$ the accretion rate is also independent of the particle number $N$.
%%Fig.5%%%%%%%%%%%%%%%%%%%
%%%%%%%%%%%%%%%%%%%%%%%

With the choosen parameters for the AD and the CSC the growth timescale of the SMBH by accretion of stars is of the order of the half-mass relaxation time $t_\mathrm{rx}$ of the CSC, which is long compared to the Eddington accretion timescale of $\approx 50$\,Myr (see Table 1). Therefore our model applies to quiesent galactic nuclei. The accretion rate can be converted to a maximum luminosity by adopting  $L_\mathrm{max}=0.1\dot{M}_\mathrm{bh}c^2$. This upper limit is for all SMBH masses in the range of $2 \dots 50 \times 10^7\Lsol$ (last column of Table 1).

\subsection{Energy dissipation} \label{sec-dissipation}

{ Our measurement of the {\em global} normalized dissipative time scale $\eta$ 
in the simulations uses 
the definition as given in Eq.~\ref{eq-eta2}.}

In a stationary state the time scale of transport of stars through the AD towards the black hole is dominated by
the outer edge of the disk, where the dissipation time scale is longest. Therefore, a measurement
of the total dissipated energy in our system should not depend on the choice of $r_\mathrm{acc}$,
where we actually remove the stars and add their mass to the central SMBH,
{ as it has been shown by Eq.~\ref{eq-nu}}.
%%Fig.6%%%%%%%%%%%%%%%%%%%
\begin{figure}
\includegraphics[width=0.48\textwidth,angle=0]{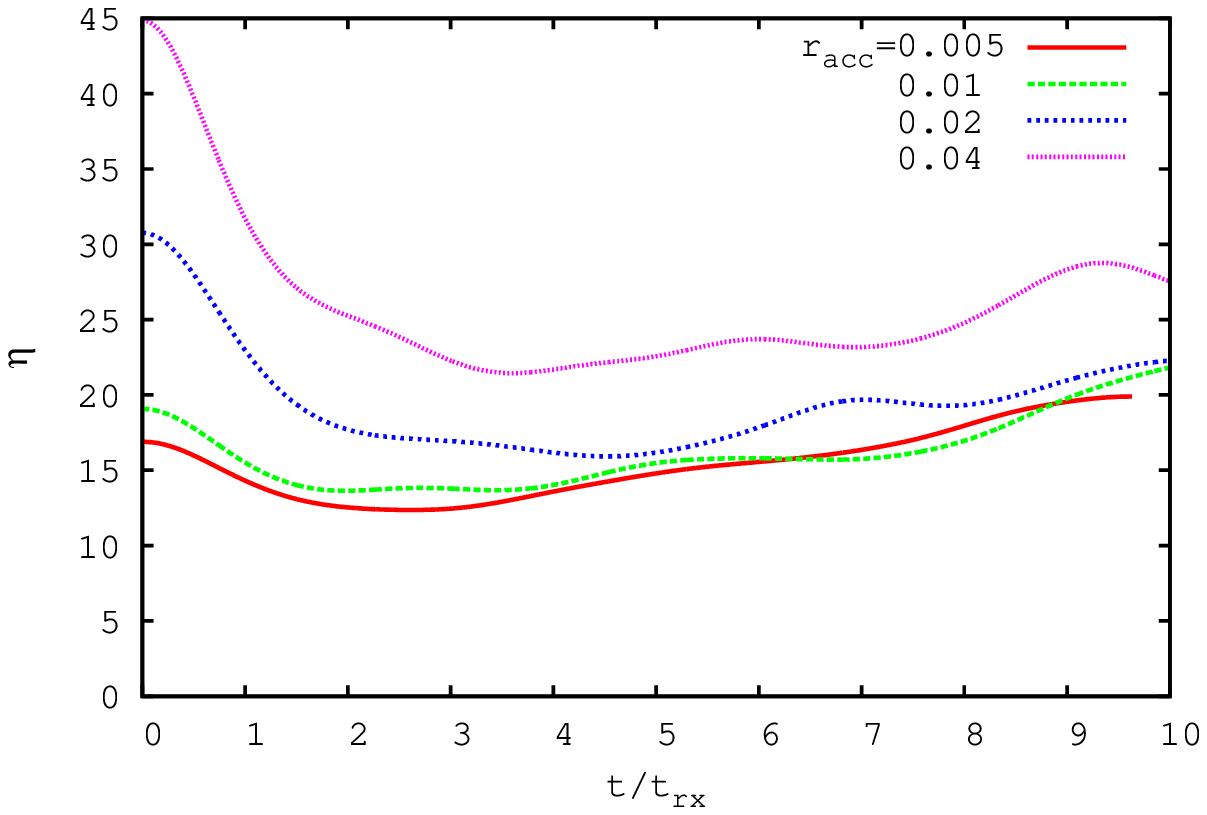}\\
\includegraphics[width=0.48\textwidth,angle=0]{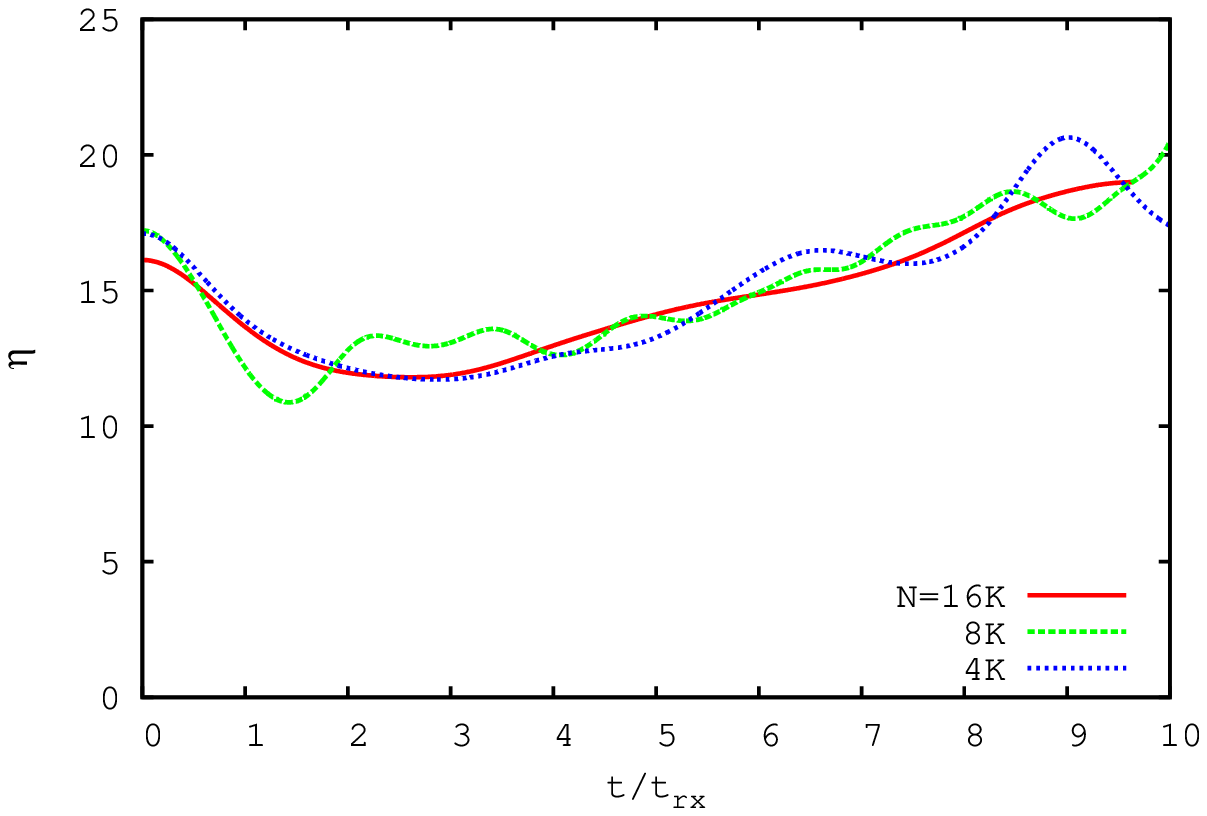}
\caption{Normalized dissipation timescale $\eta$ (Eq. \ref{eq-eta2}) for
$N=16k$ and  different $r_\mathrm{acc}$ (top panel) and for
$r_\mathrm{acc}=0.005$ and different $N$ (bottom panel).}
\label{figdotEsd}
\end{figure}
%%%%%%%%%%%%%%%%%%%%%%%
The normalized dissipation timescale $\eta$ is a measure of the dissipated energy.
It is determined by the numerical evaluation of $E_\mathrm{kin}(t)$ and
${\dot E}^\mathrm{(sd)}(t)$, the latter via smoothing of
the cumulative function $E^\mathrm{(sd)}(t)$ to derive the slope. 
After an initial adaption phase $\eta$ is approximately constant
(Fig.~\ref{figdotEsd}) showing the constant energy flow in a stationary BW cusp.
As seen in Fig.~\ref{figdotEsd}, the normalized dissipation timescale is
independent of $N$ and depends only weakly on $r_\mathrm{acc}$. 
The long-term trend of increasing $\eta$ is due to the evolution of the
SMBH and the CSC by the accretion of stars. The SMBH mass
is increasing and the mass of the CSC is decreasing leading to
an increasing
$\mu_\mathrm{bh}$ and virial factor $\xi_\mathrm{k}$, and a decreasing total
cross section $Q_\mathrm{tot}$ and efficiency $P_\mathrm{d}$ with time.
%{\bf new}
By comparing the dissipation timescale $\eta$ shown in Fig.~\ref{figdotEsd} and the accretion timescale $\nu$ shown in
Fig.~\ref{figNoverNdot} we observe that the relation derived in Eq. \ref{eq-nu} is approximately fulfilled (taking into account the growth of the black hole and the decreasing CSC mass for $\mu_\mathrm{bh}$).

%\underbar{ FIGURE 7 AND THE DISCUSSION OF IT HAS BEEN TAKEN AWAY COMPLETELY.}

\section{Discussion and Conclusions} \label{sec-conclusion}

In galactic nuclei super-massive black holes coexist with a dense stellar cluster;
galaxy mergers and the quasar phenomenon indicate that at least for some time
there should be also large amounts of interstellar gas present in the nuclear
regions around the black holes. In this paper we have examined the interaction
and co-evolution of a dense star cluster surrounding a star-accreting super-massive
black hole with an assumed central gaseous disk. Interactions of such disks with
the surrounding dense star clusters have been proposed as source of gas supply to the central
disk \citep{MiraldaK:05}, as agents to enhance the tidal star accretion rate \citep{VilkoviskiC:02},
and to cause feedback on the orbital parameters of stars \citep{Rauch:95},
including a modification of sources of gravitational waves \citep{Rauch:99}.

Our model is the first self-consistent long-term simulation of a dense star cluster,
surrounding a star-accreting super-massive black hole, and subjecting the stars
to the { dissipative} forces from a resolved central gaseous disk. We resolve
effects of two-body relaxation, dissipation of stellar kinetic energy in the
disk and star accretion onto the central black hole in a numerical study
based on direct high-accuracy simulations. Since star-by-star modeling of a
galactic nucleus down to the realistic tidal radius is not yet possible,
despite of the modern GPU hardware used for simulations, a careful scaling
analysis is presented as a function of the particle number in the simulations and
of an assumed star accretion radius, to allow conclusions for the real astrophysical
situation in galactic nuclei. But our model still has a number of serious
drawbacks. Firstly,
it assumes a stationary accretion disk, so energetic feedback to the disk structure
by star-disk interactions is neglected, secondly, the physics of
star-gas interactions
is modeled approximately. Finally, there is no distinction between
properties of 
different stars (main sequence, giants, remnants), but rather a single stellar
species with solar radius is assumed. In that sense our study should be considered as
a pathfinder and exploratory.

This investigation is a direct continuation of a semi-analytic study by
\citet{VilkoviskiC:02}, extending it by a more detailed and numerical study
of the stellar dynamical evolution of the central stellar cluster.
Our paper uses a numerical approach based on direct $N$-body
simulations, including particle-particle forces as well as a { dissipative}
force in the disk. Here we resolve the dissipation of stellar
kinetic energy along the stellar paths in the vertically extended accretion disk.
Particle numbers in our simulations and an adopted star
accretion radius (onto the central super-massive black hole) are used as
free parameters in our model, while for other
important parameters of the problem fiducial values are assumed. These are e.g.
the initial super-massive black hole mass (10 percent of 
the initial central stellar cluster mass), the gaseous accretion disk mass (10 percent of the initial 
black hole mass), and the outer radius of the accretion disk (set equal to the black hole
gravitational influence radius). Finally, all star particles are equal in the
simulation, and their total effective cross section is used as a parameter to
obtain the physically correct ratio of dissipation to relaxation time. 

We show that the accretion rate of stars onto the super-massive black hole is strongly enhanced 
by the { dissipative} force of the accretion disk (a factor of four with our parameters). The accretion 
process is determined by an equilibrium of diffusion by two-body encounters and 
energy loss by the { dissipative} force. Consistently there is also an energy deposition in
the central accretion disk; we find that most stars accreted or trapped in the
disk are quickly accreted also to the black hole, because there is no stable
co-rotation of the stars with the disk. Our results are robust with respect to
variation of particle number and adopted accretion radius, 
therefore they should hold under realistic conditions in galactic nuclei.
Our star accretion rate does not depend strongly on the
adopted star accretion radius; it supports the idea of \citet{VilkoviskiC:02}, that there
is a stationary flow of stars within the disk towards the central black hole,
which is determined by an equilibrium between dissipation and relaxation time.
In spite of the enhanced number of stars accreted through the disk, we still find
that there is a Bahcall-Wolf central density cusp present in the system, which is
not significantly perturbed.

 Central densities in star clusters near super-massive black holes can
reach
 $10^8 \Msol\,\mathrm{pc}^{-3}$ or more. At such high stellar densities direct, disruptive
stellar collisions may produce gas in the gravitational potential well of
 the black hole. The gas production rate could be larger than the one
obtained
 from tidal disruption of stars \citep{SpitzerS:66,SpitzerS:67}; see also \citet{Begelman:78,FrankR:76}, and numerical models in e.g. 
\citet{FreitagB:02,FreitagGR:06,FreitagRB:06}. But there is little
doubt that a large fraction of this gas is finally accreted to the SMBH;
some fraction of it though
 may escape. The same is true for the gas produced by tidal accretion. Our
models do not yet resolve the very central regions of the star cluster,
where
 stellar collisions may occur predominantly. 
We anyway treat the accretion radius as a free parameter used for scaling
studies; it is here usually large compared to the astrophysically defined
tidal
 radius, where stars are destroyed by tidal forces. Our results should not
depend strongly on whether the stars inside $r_\mathrm{acc}$ are finally disrupted by
tidal forces or destroyed by stellar collisions with subsequent accretion
of the gas onto the SMBH. Less is known about the
effect of
 induced stellar collisions due to low relative velocities of stars in the
disk
 (due to dissipation). We will study the effects of direct stellar
collisions
in future work.

% Rainer's suggestion, taken from abstract:

Direct stellar collisions produce another source of gas deep in the
gravitational well of the central supermassive black hole. As in the
case of tidal disruptions of stars, we do not know exactly how much
mass is accreted to the SMBH, and how much is ejected due to
magnetic fields (jets) and radiation pressure near it \citep[see e.g.
one improved model by][]{Kasen:10}. Our assumption
to add 100\% of the material from tidal accretion to the black hole
clearly is an upper limit, the real growth rate may be lower due to
some mass loss in the process, also for stellar collisions. 
However, even in our case with possibly overestimated accretion rates,
assuming a typical value of 10\% mass to energy conversion, the luminosity
obtained from our mass accretion rates are 
% Andreas additions
of the order of $10^8\Lsol$, much smaller than the
%smaller or just equal to the
Eddington luminosity. Therefore our results are applicable to quiescent
galactic nuclei, not to quasars or AGN in their most active phase, where
high mass accretion rates, feedback and outflows are driven for example
by gas inflow due to galaxy mergers.
%leading temporarily to super-Eddington accretion rates.}

%{\bf
%We have investigated the enhancement of the accretion rate of stars onto the SMBH in a 
%stationary mode of the galactic nucleus. The accretion rate scale with the dissipation 
%time scale and the growth time scale of the SMBH by this process is large compared 
%to the time scale determined by the Eddington limit for AGNs. Therefore our results 
%can be applied to accretion events on SMBHs, their structure of the AD, and to the 
%rate of stellar collisions in more quiet galactic nuclei.
%}
%

The response of the central accretion disk to the deposition of energy and stars
is neglected in this paper. We will work on this problem in future studies. 
The black hole growth rate due to star accretion will be limited by the condition
that the accretion disk can find a new equilibrium absorbing the dissipated stellar
energy and radiating it. In the inner spherically symmetric part of the system
the Eddington limit will pose an upper limit to the star accretion rate allowed
in a stationary state \citep[see discussion by][]{MiraldaK:05}. In that sense our
accretion rates, determined by neglecting feedback on the disk, will be upper
limits. It is still possible that the process of star trapping to
the disk and star and gas accretion to the central black hole is quasi-periodic and
highly non-stationary (as suggested by the ubiquitous time variability of radiation
from active galactic nuclei).

\section*{Acknowledgments}

This work has been made possible by Volkswagen Foundation
under the project ``STARDISK -- Simulating Dense Star-Gas Systems in Galactic Nuclei
using special hardware''  (I/81 396), 
for supporting collaboration with science and education in Kazakhstan. We
acknowledge personal support for DY and MM and support for German-Kazakhstan
exchange and cooperation for our entire author team. Finally, GRAPE and
GPU hardware at Fesenkov Astrophysical Institute in Kazakhstan used for this
work has been supported by the STARDISK project.

Simulations were also performed on the GRACE supercomputer
(grants I/80 041-043 and I/81 396 of the Volkswagen Foundation
and 823.219-439/30 and /36 of the Ministry of Science, Research and
the Arts of Baden-W\"urttemberg) and the special supercomputer Laohu at the 
High Performance Computing Center at National Astronomical Observatories, funded by 
the Ministry of Finance of China under 
the grant ZDYZ2008-2, has been used.
RS and PB acknowledge support by the Chinese Academy of Sciences through 
Grant Number 2009S1-5 (The Silk Road Project). 

%Computing Resources provided by 
%H\"ochstleistungsrechenzentrum Stuttgart (HLRS) 
%in the context of Baden-W\"urttemberg Grid (bwgrid) are acknowledged.

%   The special supercomputer at the Center of Information 
%and Computing at National Astronomical Observatories, 
%Chinese Academy of Sciences, funded by Ministry of Finance
%of People's Republic of China under the grant ZDYZ2008-2, 
%has been used.

PB acknowledges the special support by the NASU under 
the Main Astronomical Observatory GRID/GPU computing
cluster project. PB's studies are also partially supported 
by the program Cosmomicrophysics of NASU. A careful
reading of the preprint by Luca Naso was very helpful.
We thank the anonymous referee for an intensive discussion which led to major clarifications of the
methods and approximations
 used.

\end{document}